# Hole-Doped $M_4$SiTe$_4$ ($M$ = Ta, Nb) as an Efficient *p*-Type Thermoelectric Material for Low-Temperature Applications

Yoshihiko Okamoto,[1,a)] Yuma Yoshikawa,[1] Taichi Wada,[1] and Koshi Takenaka[1]

[1]*Department of Applied Physics, Nagoya University, Nagoya 464-8603, Japan*

Solid-state thermoelectric cooling is expected to be widely used in various cryogenic applications such as local cooling of superconducting devices. At present, however, thermoelectric cooling using *p*- and *n*-type Bi$_2$Te$_3$-based materials has been put to practical use only at room temperature. Recently, $M_4$SiTe$_4$ ($M$ = Ta, Nb) has been found to show excellent *n*-type thermoelectric properties down to 50 K. This paper reports on the synthesis of high-performance *p*-type $M_4$SiTe$_4$ by Ti doping, which can be combined with *n*-type $M_4$SiTe$_4$ in a cooling device at low temperatures. The thermoelectric power factor of *p*-type $M_4$SiTe$_4$ reaches a maximum value of approximately 60 μW cm$^{-1}$ K$^{-2}$ at 210 K and exceeds the practical level in a wide temperature range of 130–270 K. A finite temperature drop by Peltier cooling was also achieved in a cooling device made of *p*- and *n*-type Ta$_4$SiTe$_4$ whisker crystals. These results clearly indicate that $M_4$SiTe$_4$ is promising to realize a practical thermoelectric cooler for use at low temperatures, which are not covered by Bi$_2$Te$_3$-based materials.

Thermoelectric conversion is a promising all-solid-state technology that could contribute to solving global energy and environmental problems through local cooling of various devices and waste heat power generation streams. At present, the Peltier cooler using Bi$_2$Te$_3$-based materials is the only one commercially available and is used for cooling of infrared sensors and temperature control of laser diodes at around room temperature. Development of new *p*- and *n*-type materials that show high thermoelectric performance at low temperatures offers an avenue for realizing various applications, such as local cooling of cryogenic electronic devices and power generation using the cold energy of liquefied natural gas. In recent years, the development of high-temperature thermoelectric materials has been remarkable in contrast to the low-temperature materials. PbTe with hierarchical architectures, AgPb$_m$SbTe$_{2+m}$, and SnSe have been reported to show very low thermal conductivity, yielding a large dimensionless figure of merit *ZT* exceeding 2 at high temperature.[1-4] On the other hand, high performance at low temperature has been mainly achieved in studies on superlattice thin films or nanowires, which cannot be used in bulk devices.[5,6] A limited number of materials such as *n*-type Bi$_{1-x}$Sb$_x$ and *p*-type CsBi$_4$Te$_6$ and ZrTe$_5$ were reported as candidate materials for low-temperature applications, and only the *n*- or *p*-type of each shows high performance.[7-11] Discovering promising materials that show high performance in both *n*- and *p*-types at low temperatures is an important step toward the low-temperature application of thermoelectric conversion.

Recently, it was found that the one-dimensional telluride Ta$_4$SiTe$_4$ and its substituted compounds showed high *n*-type thermoelectric performance at low temperature.[12] The samples obtained have a whisker shape that is typically several mm long and several μm in diameter, reflecting the strongly one-dimensional crystal structure comprising Ta$_4$SiTe$_4$ chains.[13,14] The electrical resistivity (ρ) and thermoelectric power (*S*) measured along the Ta$_4$SiTe$_4$ chains indicate that the power factor ($P = S^2/\rho$) of Mo-doped Ta$_4$SiTe$_4$ whiskers is significantly larger than the room-temperature value of Bi$_2$Te$_3$-based materials over a wide temperature range of 50–300 K.[12] Undoped Ta$_4$SiTe$_4$ shows a very large and negative thermoelectric power of |*S*| ~ 400 μV K$^{-1}$ at 100–200 K, while maintaining a small ρ of ~ 2 mΩ cm. These *S* and ρ yield $P$ = 80 μW cm$^{-1}$ K$^{-2}$ at the optimum temperature of ~130 K, which is almost twice as large as those of Bi$_2$Te$_3$-based materials at room temperature. This *P* is strongly enhanced by Mo doping. (Ta$_{1-x}$Mo$_x$)$_4$SiTe$_4$ with $x$ = 0.001–0.002 shows $P$ = 170 μW cm$^{-1}$ K$^{-2}$ at 220–280 K. Mo-doped Nb$_4$SiTe$_4$, which is the 4*d* analogue of Ta$_4$SiTe$_4$, also showed a large *P* with the maximum value of 70 μW cm$^{-1}$ K$^{-2}$ at 230–300 K.[15]

However, Ta$_4$SiTe$_4$, Nb$_4$SiTe$_4$, and their substituted compounds reported thus far show *n*-type thermoelectric properties. Thermoelectric devices always consist of a closed circuit of *n*- and *p*-type materials and are required to stand up in tough environments such as repeated temperature cycling and large temperature gradients. Considering the thermal expansion and mechanical properties of materials, a device should consist of *n*- and *p*-type materials based on one compound. This paper reports the synthesis of high-

_______________________
a) Electronic mail: yokamoto@nuap.nagoya-u.ac.jp

performance p-type $Ta_4SiTe_4$ and $Nb_4SiTe_4$. Ti-doped $Ta_4SiTe_4$ and $Nb_4SiTe_4$ whisker crystals show a positive $S$ in a wide temperature range. $P$ of them shows a maximum value of approximately 60 $\mu$W cm$^{-1}$ K$^{-2}$ at 210 K and exceeds that of $Bi_2Te_3$-based materials in a wide temperature range from 130 to 270 K. This high performance is found in the temperature range where Ti atoms work well as an acceptor. A prototype device was also constructed by combining n- and p-type $Ta_4SiTe_4$ whisker crystals prepared by Mo and Ti doping, respectively, and a finite temperature drop by Peltier cooling was observed.

Whisker crystals of $(Ta_{1-x}Ti_x)_4SiTe_4$ and $(Nb_{1-x}Ti_x)_4SiTe_4$ ($x \leq 0.05$) were synthesized by crystal growth in a vapor phase. A stoichiometric amount of elemental powders and 100% excess of Si powder were mixed and sealed in an evacuated quartz tube with 20 mg of $TeCl_4$ powder. The tube was heated to and kept at 873 K for 24 h and at 1423 K for 96 h. The furnace was then allowed to cool to room temperature. The obtained whisker crystals are shown in the insets of Fig. 1(b) and Fig. 2(b). Majority of the whisker crystals has a diameter from 1 to 10 $\mu$m and a length from 2 to 10 mm. Relatively thick and long whisker crystals in them were used for transport measurements. Since the scanning electron microscopy observation suggested that the whisker crystals have a cylindrical shape, we evaluated the cross section of a whisker crystal for the transport measurements using the diameter measured by an optical microscope VHX-1000 (KEYENCE) by assuming the cylindrical shape. Uncertainty in the diameter in this measurement is up to 10%, indicating that the uncertainty in the evaluated cross section or electrical resistivity is up to 20%, which has little effect on the discussion in this study. Sample characterization was performed by powder X-ray diffraction analysis for pulverized whisker crystals with Cu K$\alpha$ radiation at room temperature using a RINT-2100 diffractometer (RIGAKU). In this study, nominal compositions are used to represent the chemical compositions of the whisker crystals. The electrical resistivity and thermoelectric power measurements between 5 and 300 K were performed using a Physical Property Measurement System (Quantum Design).

A thermoelectric device consisting of three pairs of bundles of $(Ta_{0.95}Ti_{0.05})_4SiTe_4$ (p-type) and $(Ta_{0.999}Mo_{0.001})_4SiTe_4$ (n-type) whisker crystals connected in series was prepared. The temperature difference between the cooling part and the heatsink (sapphire) was measured using a thermocouple, when an electric current is applied to the device.

Figures 1(a) and 1(b) show the temperature dependences of $S$ and $\rho$ whisker crystals of $(Ta_{1-x}Ti_x)_4SiTe_4$ ($x = 0.001$–0.05), respectively, measured along the $c$ axis, i.e., parallel to the chain direction. The data for undoped $Ta_4SiTe_4$ are also shown as a reference.[12] Although the undoped $Ta_4SiTe_4$ shows negative $S$ below room temperature, only 0.1

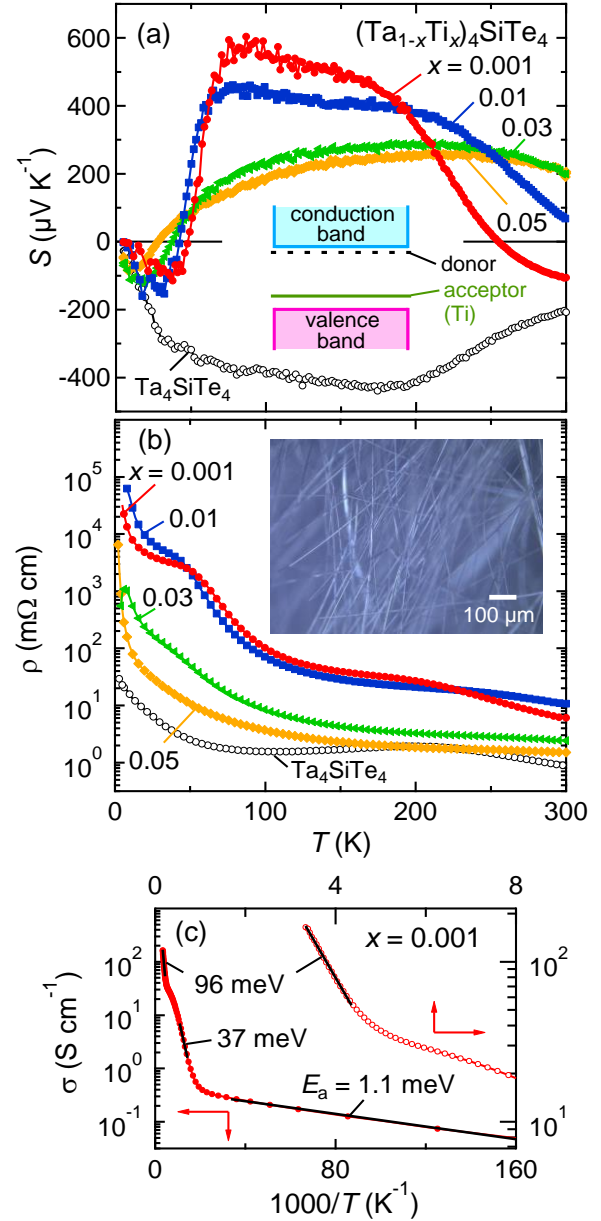

FIG. 1. Temperature dependences of (a) thermoelectric power and (b) electrical resistivity of whisker crystals of $(Ta_{1-x}Ti_x)_4SiTe_4$ (0.001 $\leq x \leq$ 0.05) measured along the $c$ axis are shown. Data for $Ta_4SiTe_4$ are also shown as a reference.[12] (c) An Arrhenius plot of electrical conductivity at $x = 0.001$. The open circles indicate an enlarged view of the high temperature region. Solid lines represent the results of linear fits to 230–300 K, 70–92 K, and 6–30 K data. The inset in (a) shows a schematic energy diagram of Ti-doped $Ta_4SiTe_4$. The broken and solid lines indicate the donor and acceptor levels, respectively. The inset in (b) shows an optical microscope image of whisker crystals of Ti-doped $Ta_4SiTe_4$. The bar in the image indicates 100 $\mu$m.

% Ti doping results in positive $S$ between 50 and 250 K. $S$ of the $x = 0.001$ sample is negative below 50 K, strongly increases at around 50 K with increasing temperature, and then shows a maximum value of $S = 560$ $\mu$V K$^{-1}$ at 90 K. This $S$ far exceeds the maximum $|S|$ of ~400 μV K$^{-1}$ in undoped $Ta_4SiTe_4$, indicating that this system is also promising as a $p$-type material. Further increases in temperature lead to gradual and then strong decreases in $S$, resulting in negative $S$ once again above 250 K. On the other hand, ρ of the $x = 0.001$ sample is more than an order of magnitude larger than that of the undoped one. ρ = 30–100 mΩ cm in the temperature range of 80–200 K, where large and positive $S$ is realized, is too large for a thermoelectric material. ρ of the $x = 0.001$ sample shows complex temperature dependence, with humps at around 50 and 200 K. These anomalies correspond to the changes in the temperature dependence of $S$, as shown in Fig. 1, suggesting that these anomalies are intrinsic to this system, which will be discussed with the results on $Nb_4SiTe_4$.

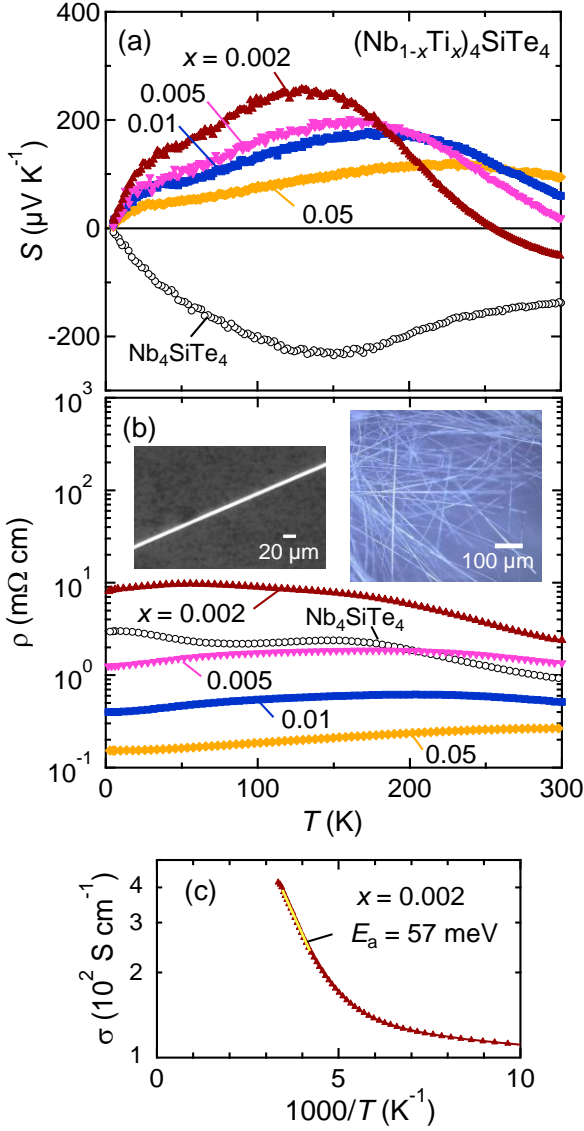

FIG. 2. Temperature dependences of (a) thermoelectric power and (b) electrical resistivity of whisker crystals of $(Nb_{1-x}Ti_x)_4SiTe_4$ ($0.002 \leq x \leq 0.05$) measured along the $c$ axis are shown. Data for $Nb_4SiTe_4$ are also shown as a reference.[15] (c) An Arrhenius plot of the electrical conductivity at $x = 0.002$. The solid line represents the result of a linear fit to 238–290 K data. The insets in (b) shows scanning electron microscope (left) and optical microscope (right) images of whisker crystals of Ti-doped $Nb_4SiTe_4$. The bar in each image indicates 20 and 100 μm, respectively.

With increasing Ti content, $S$ and ρ of the Ti-doped $Ta_4SiTe_4$ showed systematic changes. The temperature range with positive $S$ is extended to both lower and higher temperatures, accompanied by a decrease in the maximum value of $S$. These doping dependences are similar to those of Mo-doped $Ta_4SiTe_4$, except for the sign changes.[12] The $x = 0.05$ sample shows $S > 200$ μV K$^{-1}$ in the wide temperature range from 110 to over 300 K, with the maximum value of $S = 250$ μV K$^{-1}$ at 220 K. This $S$ is large enough for a thermoelectric material and comparable to those of $Bi_2Te_3$-based materials at room temperature. ρ of $(Ta_{1-x}Ti_x)_4SiTe_4$ decreases with increasing $x$, consistent with the expectation that Ti substitution to the Ta site is hole doping, which increases the number of hole carriers. The $x = 0.05$ sample shows small ρ of 2 mΩ cm at 220 K, where $S$ shows the maximum value, although the metallic behavior with $d\rho/dT > 0$ does not appear to be different from that of the $n$-type $Ta_4SiTe_4$.[12]

Figures 2(a) and 2(b) show the temperature dependences of $S$ and ρ, respectively, for whisker crystals of $(Nb_{1-x}Ti_x)_4SiTe_4$ ($x = 0.002$–0.05) measured along the $c$ axis. The data for undoped $Nb_4SiTe_4$ are also shown as a reference.[15] Although Ti-doped $Nb_4SiTe_4$ samples showed $p$-type behavior with positive $S$, similar to the $Ta_4SiTe_4$ case, the $S$ and ρ values of Ti-doped $Nb_4SiTe_4$ are smaller than those of the Ta compounds. The most lightly doped $x = 0.002$ sample shows a positive $S$ below 250 K with a maximum value of $S = 250$ μV K$^{-1}$ at $T_p = 130$ K, which is comparable to that of $n$-type $Nb_4SiTe_4$.[15] In contrast to Ti-doped $Ta_4SiTe_4$, Ti-doped $Nb_4SiTe_4$ shows positive $S$ down to the lowest measured temperature, which will be discussed later. With increasing $x$, the maximum value of $S$ decreases and $T_p$ is shifted to higher temperature. The $x = 0.05$ sample shows a maximum $S$ of 120 μV K$^{-1}$ at 230 K. This doping dependence is similar to that of Mo-doped $Nb_4SiTe_4$.[15] On the other hand, ρ of the $x = 0.002$ sample is much larger than that of undoped $Nb_4SiTe_4$. The value of ρ strongly decreases and the temperature range with $d\rho/dT > 0$ is extended with increasing $x$, resulting in a metallic temperature dependence of $x = 0.05$ in the entire measured temperature range.

Figure 3 shows the power factor $P$ of the whisker crystals of $(Ta_{1-x}Ti_x)_4SiTe_4$ and $(Nb_{1-x}Ti_x)_4SiTe_4$. The largest $P$ of the $p$-type $M_4SiTe_4$ is realized in $(Nb_{0.95}Ti_{0.05})_4SiTe_4$, which shows a maximum value of $P_{max} = 57$ μW cm$^{-1}$ K$^{-2}$ at the optimum temperature of 210 K. $P_{max}$ values of $(Nb_{0.99}Ti_{0.01})_4SiTe_4$ and $(Ta_{0.95}Ti_{0.05})_4SiTe_4$ are also larger than the 35 µW cm$^{-1}$ K$^{-2}$ for Bi$_2$Te$_3$-based materials.[16,17] These power factors are smaller than $P_{max}$ = 170 µW cm$^{-1}$ K$^{-2}$ in the $n$-type sample,[12] but exceed the practical level in the wide temperature range from 130 to 270 K, indicating that the $p$-type $M_4SiTe_4$ samples prepared in this study have high enough performance to be combined with the $n$-type ones in a thermoelectric device. The large power factor of $p$-type $M_4SiTe_4$ is expected to be related to its characteristic band structure with a very small band gap and strongly one-dimensional band dispersion, as in the case of $n$-type $M_4SiTe_4$.[12,15]

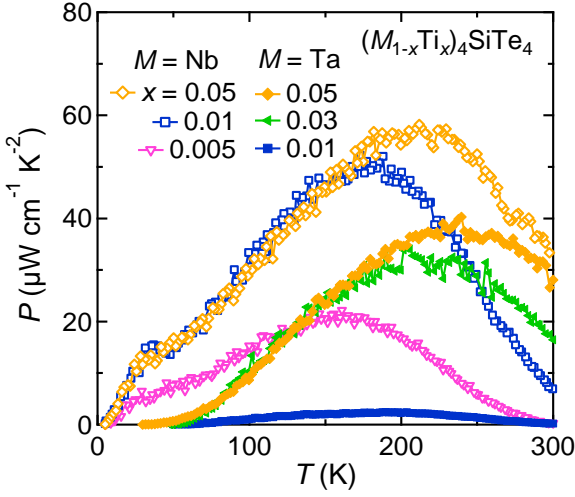

FIG. 3. Thermoelectric power factor of $(M_{1-x}Ti_x)_4SiTe_4$. Filled and open symbols represent data for $M$ = Ta and Nb, respectively. Data with $p$-type thermoelectric properties are shown.

Here we discuss the thermoelectric properties of $p$-type $M_4SiTe_4$, mainly by using experimental data for $(Ta_{0.999}Ti_{0.001})_4SiTe_4$, where the thermally activated behavior is most clearly observed. As shown in the Arrhenius plot of the electrical conductivity ($\sigma = 1/\rho$) of $(Ta_{0.999}Ti_{0.001})_4SiTe_4$ shown in Fig. 1(c), linear behavior appears in three temperature regions: $T$ < 30 K (1000/$T$ > 30 K$^{-1}$), $T$ ~ 80 K (1000/$T$ ~ 14 K$^{-1}$), and $T$ > 230 K (1000/$T$ < 4 K$^{-1}$). The activation energies of each temperature range are estimated to be $E_a$ = 1.1, 37, and 96 meV, respectively.

Thermally activated behavior at the low temperature of $T$ < 30 K is considered to be due to thermal excitation of electrons from the donor levels, which is intrinsic to Ta$_4$SiTe$_4$. Ta$_4$SiTe$_4$ is naturally electron-doped, which is probably due to the tellurium vacancy, resulting in negative $S$ in $(Ta_{1-x}Ti_x)_4SiTe_4$ at low temperature.

The linear behavior in the middle temperature range at $T$ ~ 80 K is considered to be due to thermal excitation from the valence band to the acceptor levels introduced by Ti doping. Hole carrier density is expected to be strongly increased in this temperature range, resulting in compensation of electron carriers present at lower temperatures and in large and posi-tive $S$ above this temperature. The acceptor levels are probably located at the energy corresponding to this temperature above the top of the valence band, as shown in Fig. 1(c), meaning that they do not work as an acceptor at lower temperatures. This result suggests that the position of the acceptor level introduced by hole doping is important to realize a $p$-type material that works effectively at low temperatures. For example, $S$ of Ti-doped Nb$_4$SiTe$_4$ is positive down to the lowest measured temperature, as seen in Fig. 2(a), suggesting that the acceptor levels in Nb$_4$SiTe$_4$ are located closer to the top of the valence band than those in Ta$_4$SiTe$_4$, which might reflect a smaller band gap in Nb$_4$SiTe$_4$ than in Ta$_4$SiTe$_4$, as discussed below.

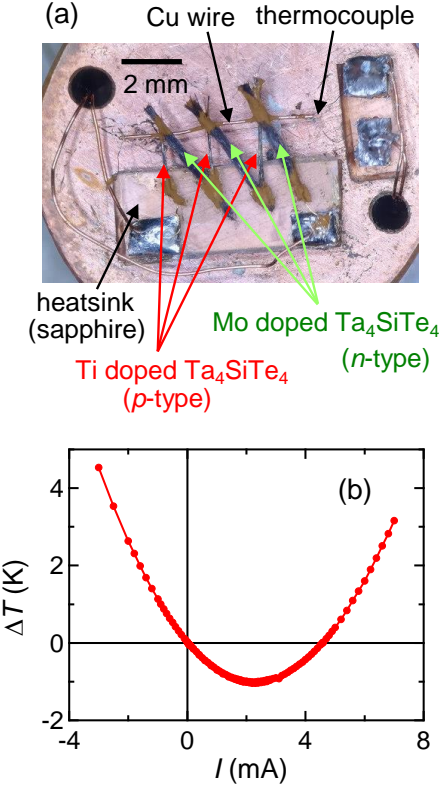

FIG. 4. Estimation of thermoelectric performance of a thermoelectric device made of Ta$_4$SiTe$_4$ whisker crystals. (a) The prepared device. (b) Temperature difference between the cooling part and the heatsink versus electrical current applied to the device at room temperature.

The other thermally activated behavior with $E_a$ = 96 meV at high temperature above 230 K might have corresponded to the thermal excitation of electrons across the band gap. This would be consistent with strong decreases in |$S$| above 200 K in the $x$ = 0.001 and 0 samples. The first-principles calculation results indicated that Ta$_4$SiTe$_4$ has a small band gap of $\Delta$ ~ 0.1 eV due to the spin–orbit coupling effect.[12] The observed behavior at high temperatures supports the presence of this band gap. Moreover, $\rho$ of Nb$_4$SiTe$_4$ shows a thermally activated behavior with smaller $E_a$ than in the Ta case. As shown in Fig. 2(c), the linear fit to the Arrhenius plot of σ of $(Nb_{0.998}Ti_{0.002})_4SiTe_4$ above 238 K yields $E_a$ = 57 meV, which is almost half the $E_a$ of $(Ta_{0.999}Ti_{0.001})_4SiTe_4$, supporting the assertion that spin−orbit coupling plays an important role in forming the band gap in $M_4SiTe_4$. These results indicate that materials with a small spin−orbit gap are promising as high-performance thermoelectric materials for low-temperature applications.

Finally, we describe preparation of a prototype thermoelectric device made of p- and n-type $Ta_4SiTe_4$ whisker crystals. Figure 4(a) shows a thermoelectric device consisting of three pairs of bundles of Ti- and Mo-doped $Ta_4SiTe_4$ whisker crystals. The Ta-based whisker crystals were used as both n- and p-type materials because they are longer and thicker than the Nb-based ones. As shown in Fig. 4(b), a temperature drop up to $|\Delta T|$ = 1 K occurred, accompanied by a quadratic relation between $\Delta T$ and the applied electric current I. This quadratic curve has a minimum at $I \neq 0$, suggesting that a Peltier heat absorption proportional to I and a Joule heat proportional to $I^2$ coexisted in the device and that the temperature in the cooling part was decreased by the former effect. The small temperature drop of 1 K might be due to small cooling power of this device consisting of the whisker crystals, which is not enough to compensate heat flow by radiation and thorough the wires such as thermocouples. We expect that the large cooling power at low temperature will be realized in the device using bulk-size $M_4SiTe_4$ materials obtained by the preparation of single crystals or unidirectionally oriented whisker crystals.

In summary, this paper described the synthesis of p-type $M_4SiTe_4$ by Ti doping and found that the Ti-doped samples show large P exceeding the practical level in a wide temperature range from 130 to 270 K. This result means that a couple of p- and n-type thermoelectric materials based on one compound were obtained, which is promising for low-temperature applications. Actually, a prototype thermoelectric device consisting of p- and n-type $Ta_4SiTe_4$ whisker crystals was prepared, and a finite temperature drop by Peltier cooling was observed, which is believed to be an important step for the practical use of $M_4SiTe_4$ in the future. It was also demonstrated that $M_4SiTe_4$ has a small band gap due to spin−orbit coupling, which was predicted in the first-principles calculation results, and high thermoelectric performance in the p-types is realized in the temperature range where the dopants effectively work as an acceptor. This might be important for discovering further low-temperature thermoelectric materials.

**ACKNOWLEDGMENTS**

We are grateful to Y. Yamakawa and A. Yamakage for helpful discussions. This work was partly supported by JSPS KAKENHI (Grant Nos. 16H03848 and 18H04314), the Asahi Glass Foundation, the Research Foundation for the Electrotechnology of Chubu, and the Inamori Foundation.